\documentclass{elsart}
\usepackage{epsf}
\def\figuresize{12cm}

\begin{document}
\begin{frontmatter}
\title{Transport in disordered interacting systems: Numerical results
   for one-dimensional spinless electrons}
\author{Michael Schreiber, Frank Epperlein, and Thomas Vojta}
\address{Institut f\"ur Physik, Technische Universit\"at, D-09107 Chemnitz, Germany}
\begin{abstract}
The combined influence of disorder and interactions on the transport properties of  electrons
in one dimension is investigated. The numerical simulations are carried out by means of the
Hartree-Fock-based diagonalization (HFD), a very efficient method to determine the low-energy
properties of a disordered many-particle system. We find that the conductance of a strongly 
localized system  can become considerably enhanced by the interactions. The enhancement 
for long-range interactions is significantly larger than for short-range interactions.  In contrast,
the conductance  of  weakly localized systems becomes suppressed by the interactions.
\end{abstract}
\end{frontmatter}

The transport properties of disordered electrons have been a subject of continuous
interest within the last 4 decades. In 1958 Anderson pointed out \cite{anderson}
that the electronic (single-particle) wave function may become localized
in space for sufficiently strong disorder. For one-dimensional systems it was 
later proved rigorously that all states are exponentially localized for arbitrary
finite disorder \cite{ishii,erdoes,froehlich}. Further investigations of {\em non-interacting}
disordered electrons led to the development of the scaling theory of localization
\cite{aalr,kramrev93}. 
It predicts that in the absence of a magnetic field or spin-orbit
coupling all states are localized not only in one but also in two dimensions. 
Thus a metallic state does not
exist in these dimensions. In contrast, in three dimensions states are extended for 
weak disorder while they are localized for sufficiently strong disorder. This gives rise to a 
disorder-driven metal-insulator transition (MIT) at a certain value of disorder strength.

Later also the influence of electron-electron interactions on the transport properties of disordered
electrons was investigated intensively by means of  many-body perturbation theory \cite{altshuler}, 
scaling theory \cite{finkelstein}, and
the renormalization group (for reviews see. e.g. \cite{leerev85,polrev85,belitzrev94}).  
This led to a qualitative analysis of 
the MIT and the identification of the different universality classes. 
One of the main results is that the lower critical dimension of the MIT is $d_c^-=2$ as it is
for non-interacting electrons \cite{belitzrev94}.
Therefore it came as a surprise when measurements \cite{2DMIT} 
on Si-MOSFETs revealed indications of a MIT in 2D. Since these experiments
are carried out at low electron density where the Coulomb interaction is particularly strong 
compared to the Fermi energy,
interaction effects are the most likely reason for this phenomenon. A complete understanding
has, however, not yet been obtained. There have been attempts to explain the experiments
based on the perturbative
renormalization group \cite{runaway}, non-perturbative effects 
\cite{nonperturb}, or  the transition being a superconductor-insulator transition rather than a MIT
\cite{SIT}.

In view of all this it seems to be important to investigate the problem of interacting disordered
electrons not only in the perturbative regime (of weak disorder and interactions) but also for
strong disorder or/and interactions. Recently, we investigated \cite{HFD} the transport properties
of two-dimensional disordered interacting electrons. We found that weak interactions enhance the
conductance in the strongly localized regime while they reduce the conductance 
in the case of weaker disorder. In contrast,
sufficiently strong interactions always reduce the conductance.

In this paper we extend this study to the case of one dimension. We report numerical results 
for the conductance of a simple model system of interacting disordered electrons, viz. spinless fermions in 
a random potential interacting via Coulomb or short-range interactions. Our results cover the entire 
parameter range from weak disorder and interactions to strong disorder and interactions.
The model, a one-dimensional version of the quantum Coulomb glass model 
\cite{efros95,talamantes96,epper_hf,epper_exact},  is defined on a ring (using
periodic boundary conditions) of $L$ 
sites occupied by $N=K L $ electrons ($0\!<\!K\!<\!1$). To ensure charge neutrality
each lattice site carries a compensating positive charge of  $Ke$. The Hamiltonian
is given by
\begin{equation}
H =  -t  \sum_{\langle ij\rangle} (c_i^\dagger c_j + c_j^\dagger c_i) +
       \sum_i \varphi_i  n_i + \frac{1}{2}\sum_{i\not=j}(n_i-K)(n_j-K)U_{ij}
\label{eq:Hamiltonian}
\end{equation}
where $c_i^\dagger$ and $c_i$ are the electron creation and annihilation operators
at site $i$, respectively,  and $\langle ij \rangle$ denotes all pairs of nearest 
neighbor sites.
$t$ gives the strength of the hopping term and $n_i$ is the occupation number of site $i$. 
For a correct description of the insulating phase the Coulomb 
interaction between the electrons is kept long-ranged,
$U_{ij} = U/r_{ij}$, since screening breaks down in the insulator
($r_{ij}$ is measured in units of the lattice constant). 
For comparison we also investigate the case of nearest neighbor interaction
of strength $U$.
The random potential values $\varphi_i$ are chosen 
independently from a box distribution of width $2 W$ and zero mean.
(In the following we always set $W=1$.)
Two important limiting cases of the quantum Coulomb glass are the Anderson model of
localization (for $U_{ij}=0$) and the classical Coulomb glass (for $t=0$).

The simulation of disordered quantum many-particle systems is numerically very costly
since the size of the Hilbert space grows exponentially with system size and
since many disorder configurations have to be considered to obtain typical values or
distribution functions of observables.
We have recently developed the Hartree-Fock based diagonalization (HFD) method \cite{HFD}
for the simulation of disordered quantum many-particle systems.
This method which is based on the idea of the configuration interaction approach
\cite{fulde} adapted to disordered lattice models is
very efficient in calculating low-energy properties in any spatial dimension and for 
short-range as well as long-range interactions.
It consists of 3 steps: (i) solve the Hartree-Fock (HF) approximation of the Hamiltonian, 
(ii) use a Monte-Carlo algorithm to find the low-energy many-particle HF states, 
(iii) diagonalize the Hamiltonian in the basis formed by these states. 
The efficiency of the HFD method is due to the fact that the HF basis states are 
comparatively close in character to the exact eigenstates in the entire
parameter space \cite{othermethods}. 
Thus it is sufficient to keep only a small fraction of the Hilbert 
space in order to obtain low-energy  quantities with an accuracy comparable to that of exact 
diagonalization. 
For the present study we have simulated rings with 24 sites and 12 electrons. For this
size we found it sufficient to keep 500 to 1000  (out of $2704156$) basis states.

The conductance is calculated from the Kubo-Greenwood 
formula \cite{kubo_greenwood} which connects the conductance with the
current-current correlation function in the ground state. Using the spectral
representation of the correlation function the real (dissipative) part 
of the conductance (in units of the quantum conductance $e^2/h$)
is obtained as 
\begin{equation}
 \Re ~ G^{xx}(\omega) = \frac {2 \pi^2}   {L \omega} \sum_{\nu} |\langle 0 | j^x|\nu \rangle |^2 
     \delta(\omega+E_0-E_{\nu})
\label{eq:kubo}
\end{equation}
where $j^x$ is the $x$ component of the current operator and $\nu$ denotes the eigenstates
of the Hamiltonian.  The finite life time $\tau$ of the eigenstates in a real d.c.\ transport experiment
(where the system is not isolated but connected 
to contacts and leads) results in an inhomogeneous broadening $\gamma = 1/\tau$
of the $\delta$ functions in the Kubo-Greenwood formula \cite{datta}. Here we have
chosen $\gamma=0.05$ which is of the order of the single-particle level spacing.

\begin{figure}
  \epsfxsize=\figuresize
  \centerline{\epsffile{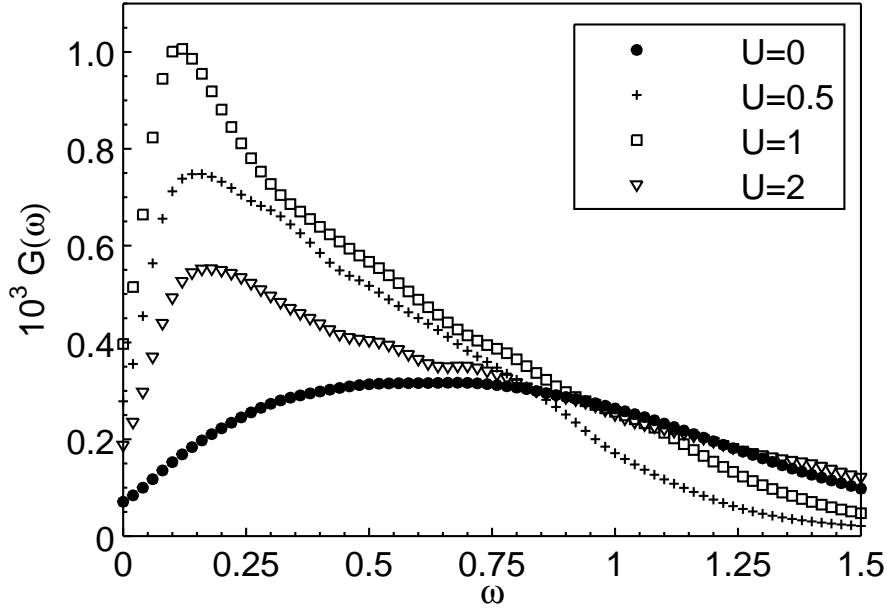}}
  \caption{Conductance $G$ as a function of frequency $\omega$,
              $W=1$,
              $t=0.03$, $\gamma=0.05$.  The truncation of the Hilbert space
              to 500 basis states restricts the validity of these data to frequencies $\omega<0.75$.}
  \label{fig:sig_a}
\end{figure}

\begin{figure}
  \epsfxsize=\figuresize
  \centerline{\epsffile{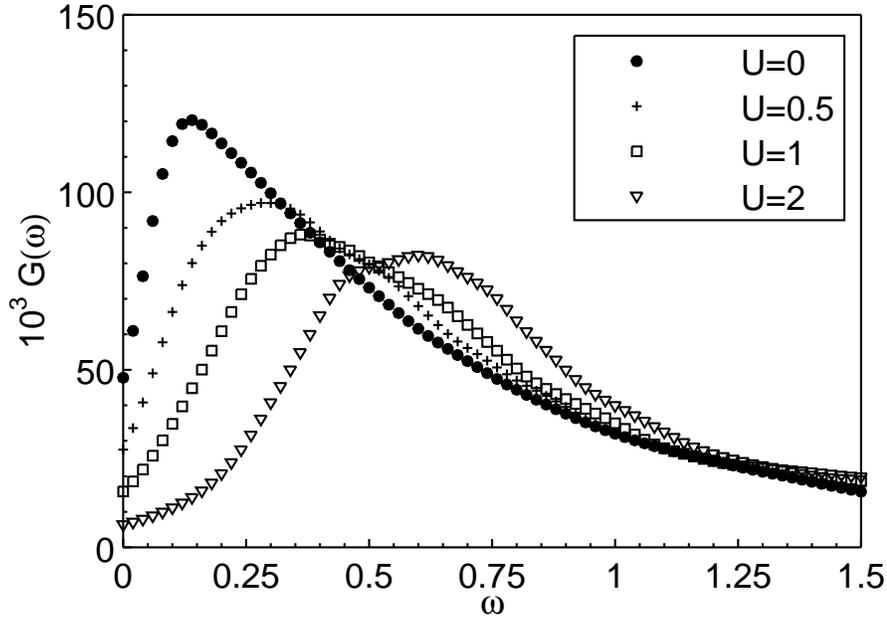}}
  \caption{Same as Fig.\ \protect\ref{fig:sig_a} but for $t=0.5$.}
  \label{fig:sig_b}
\end{figure}

We now present results on the dependence of the conductance on the
interaction strength.
In Figs.\ \ref{fig:sig_a} and \ref{fig:sig_b} we show the conductance as a function 
of frequency  for two sets of parameters. The data represent logarithmic averages 
over 400 disorder  configurations.
In Fig.\ \ref{fig:sig_a} the kinetic
energy is very small ($t=0.03$). Thus the system is in the highly localized regime,
as we have also estimated from the single-particle participation number 
which is smaller than 2.
Here not too strong Coulomb interactions ($U = 0.5, 1.0$) lead to
an {\em increase} of the conductance at low frequencies. If the interaction 
becomes stronger ($U=2$) the conductance finally decreases again.
The behavior is qualitatively different at higher kinetic energy 
($t=0.5$) as shown in Fig.\ \ref{fig:sig_b}. Here
the localization is much weaker (the
single-particle participation number is of the order of 10).
Already  a weak interaction ($U=0.5$) leads to a reduction of the 
low-frequency conductance
compared to non-interacting electrons. If the interaction becomes stronger ($U=2$)
the conductance decreases further.

\begin{figure}
  \epsfxsize=\figuresize
  \centerline{\epsffile{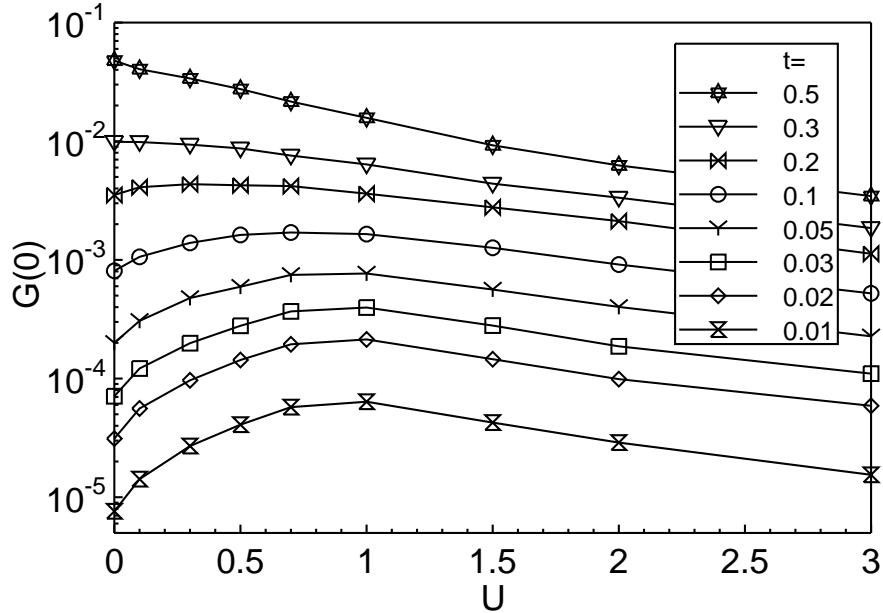}}
  \caption{d.c.\ conductance $G(0)$ as a function of interaction strength $U$
              for different kinetic energies $t$.}
  \label{fig:sigzero}
\end{figure}

We have carried out analogous calculations for kinetic energies $t=0.01, ..., 0.5$ and
interaction strengths $U=0, ..., 3$. The resulting d.c.\ conductances are presented
 in Fig.\ \ref{fig:sigzero} which is the main result of 
this paper.
It  shows that the influence of weak repulsive electron-electron interactions on the d.c.\ 
conductance is opposite in the cases of weak and strong disorder.
Sufficiently strong interactions always reduce the conductance. This is not surprising since strong
interactions will reduce charge fluctuations and in the 
limit of  infinite interaction strength the system approaches a Wigner crystal. 
In contrast, the effect of weak 
(compared to the random potential) interactions depends on the value $t$ of the kinetic energy. 
The conductance of 
strongly localized samples ($t=0.01, ..., 0.05$) becomes considerably enhanced 
by a weak Coulomb interaction. In this regime the dominant effect is the suppression
of  the localizing interference effects by electron-electron scattering events.
With increasing kinetic energy the relative enhancement
decreases as does the interaction range where the enhancement occurs. The conductance
of samples with high kinetic energies ($t \ge 0.3$) is reduced even 
by weak interactions. Here the dominant effect is the suppression of charge fluctuations
by the interactions. Overall, only the behavior at high kinetic energies (i.e. weak disorder)
is in agreement with analytical results based on the perturbative renormalization group \cite{1drg}
while the behavior for low kinetic energy (strong disorder) is qualitatively different.

\begin{figure}
  \epsfxsize=\figuresize
  \centerline{\epsffile{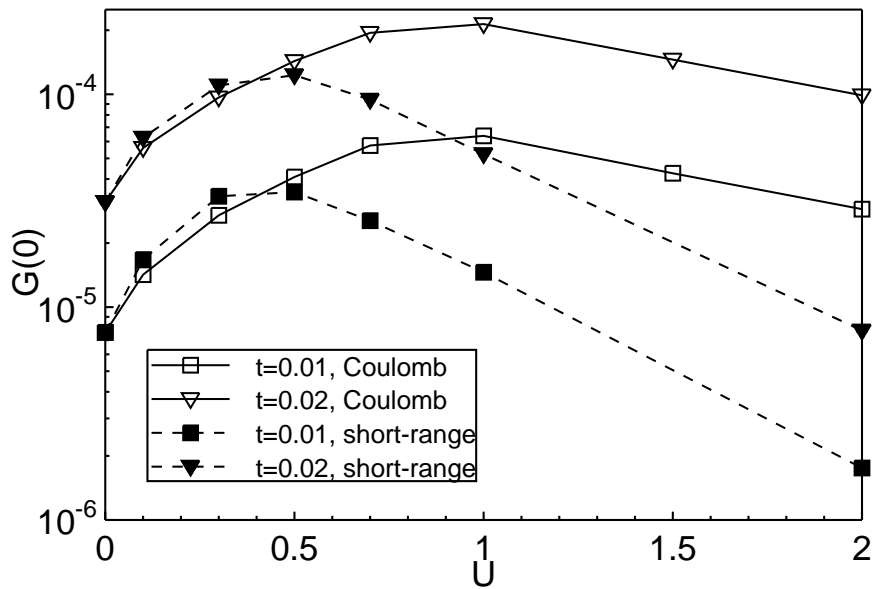}}
  \caption{d.c.\ conductance $G(0)$ as a function of interaction strength $U$
              for short- and long-range interactions.
              The data represent logarithmic averages over 400 disorder  configurations.}
  \label{fig:sr_lr}
\end{figure}

For comparison we have also investigated nearest-neighbor interactions instead of 
long-range Coulomb interactions.  In Fig.\ \ref{fig:sr_lr} we compare
the d.c.\ conductances of  systems with short- and long-range interactions in the localized
regime. 
The data show that the interaction induced enhancement of the conductance is overall 
weaker in the case of short range interactions. In particular, the maximum enhancement
as a function of interaction strength is significantly smaller (by a factor of about two) than
in the Coulomb case. Moreover, the maximum occurs for weaker interaction strength.

In the last part of this paper we want to relate our findings to the two-dimensional case \cite{HFD}
and to results in the literature. The qualitative dependence of the d.c.\ conductance on kinetic
energy and interaction is identical in one and two spatial dimensions. The interaction-induced 
enhancement in the localized regime is, however, significantly larger in the one-dimensional
systems investigated. Up to now it is not clear whether this is a true dimensionality effect
or a result of the different linear system sizes studied. In order to resolve this question a systematic 
investigation of the system size dependence is in progress \cite{method}. 
The resulting scaling behavior of the conductance with system size will also allow us to check 
for the existence of an interaction-induced MIT.
Note, however, that the recently observed MIT in 2D MOSFETs \cite{2DMIT} is not likely
to be explained by the enhancement of the conductance we found since the importance of the spin
degrees of freedom for this transition is well established experimentally \cite{magnet}.
We emphasize,
 however, that our numerical method is very easy to generalize
to electrons with spin. The fact that we find the strongest enhancement of the conductance
for very low kinetic energy also suggests that the mechanism is different from that giving
rise to an increased two-particle localization length in the problem of just two interacting
particles \cite{TIP} (where the strongest delocalization is observed for weak disorder).
Let us finally mention that the conclusions drawn in this paper are in qualitative agreement with
those of a recent DMRG study \cite{fermimott} of the phase sensitivity of the ground state energy 
for a disordered one-dimensional  model of spinless fermions with nearest-neighbor interactions
which showed that for small disorder repulsive interactions reduce the phase sensitivity 
while for large disorder  the phase sensitivity shows pronounced enhancements 
at certain values of the interaction.

To summarize, we have used the Hartree-Fock based diagonalization (HFD) method
to investigate the conductance and localization properties of disordered interacting spinless
electrons in one dimension. We have found that a weak Coulomb interaction 
can enhance the conductivity of strongly localized samples by almost one order of magnitude,
while it reduces the conductance of weakly disordered samples. If the interaction becomes stronger 
it eventually reduces the conductance also in the localized regime. The interaction
induced enhancement is larger for long-range interactions than for short-range interactions.

We acknowledge financial support by the Deutsche Forschungsgemeinschaft.


\begin{thebibliography}{9}
\bibitem{othermethods}Similar methods based on eigenstates of either the non-interacting or the
      classical problem are efficient only in a small part of the parameter space.
\end{thebibliography}
\end{document}